\documentclass[aps,prl,tightenlines,onecolumn,superscriptaddress]{revtex4}
\usepackage{amssymb}
\usepackage{color,graphicx}
\usepackage{lmodern}
\usepackage{amsmath}
\usepackage{amsbsy}
\usepackage{amsthm}
\usepackage{float}
\usepackage{bbm}
\usepackage{bm}
\usepackage{subfigure}
\usepackage[paperwidth=210mm,paperheight=297mm,centering,hmargin=1.5cm,vmargin=2cm]{geometry}

\begin{document}
\title{Validity Examination of the Dissipative Quantum Model of Olfaction}
\author{Arash Tirandaz}
\affiliation{School of Biological Sciences, Institute for Research in Fundamental Sciences (IPM), P.O. Box 19395-5531, Tehran, Iran}
\author{Farhad Taher Ghahramani}
\affiliation{School of Physics, Institute for Research in Fundamental Sciences (IPM), P.O. Box 19395-5531, Tehran, Iran}
\author{Vahid Salari}
\email[]{vahidsalari@cc.iut.ac.ir}
\affiliation{School of Physics, Institute for Research in Fundamental Sciences (IPM), P.O. Box 19395-5531, Tehran, Iran}
\affiliation{Department of Physics, Isfahan University of Technology, Isfahan P.O. Box 84156-83111, Iran}

\begin{abstract}
  The validity of the dissipative quantum model of olfaction has not been examined yet and therefore the model suffers from the lack of experimental support. Here, we generalize the model and propose a numerical analysis of the dissipative odorant-mediated inelastic electron tunneling mechanism of olfaction, to be used as a potential examination in experiments. Our analysis gives several predictions on the model such as efficiency of elastic and inelastic tunneling of electrons through odorants, sensitivity thresholds in terms of temperature and pressure, isotopic effect on sensitivity, and the chiral recognition for discrimination between the similar and different scents. Our predictions should yield new knowledge to design new experimental protocols for testing the validity of the model.
\end{abstract}

\maketitle
\section{Introduction}
Olfaction seems to be an immediate and intimate sense but surprisingly the mechanism is still not well understood. This is an important problem in its own right for both fundamental science and industry~\cite{Row,Axe,Buc,LeeK,Fara}. The olfactory system in human beings is triggered by binding the small, neutral, and volatile molecules known as odorants to specific sites on olfactory receptors (ORs) in the nasal cavity (see Figure 1). Despite considerable knowledge of structure of ORs, the detailed molecular mechanisms for discrimination between different odorants are not yet fully understood~\cite{Zar}. In 1963, Amoore conjectured that such molecular mechanism is primarily related to the {\it shape} of the odorant and accordingly it is initiated by a mutual structural fit between the odorant and ORs (i.e. {\it lock and key} model)~\cite{Amo}. In fact, the idea was motivated from the molecular mechanism of the enzyme behaviour. The model can be modified by introducing a distortion of the whole system to induce a more appropriate mutual fit (i.e. {\it hand and glove} model). A more refined demonstration of the idea requires that ORs respond to only one structural feature, such as a functional group, instead the main body of the odorant (i.e. {\it odotope} model)~\cite{Mor}. There is plenty of evidence for cases where the structure does seem important to an odorant's detection (e.g. see~\cite{Yos,Ara}). Despite the predictive power of these structure-based models, there are some evidence against them: odorants that smell similarly whilst being structurally different (e.g. benzaldehyde versus hydrogen cyanide), and odorants that smell differently whilst being structurally the same (e.g. ferrocene versus nickelocene)~\cite{Tur,Ben,Bro0}. All such shape-based models are primarily based on mechanical mechanisms.\\
\indent The quantum model of olfaction, which was firstly proposed by Dyson~\cite{Dys} and refined by Wright~\cite{Wri}, is based on the idea that the signature of scent is due to the odorant's unique vibrational spectrum not its structure. An unique scent is attributed to its unique spectrum in the same way a colour is associated to its unique frequency of light.

\begin{figure}[H]
\includegraphics[scale=0.5]{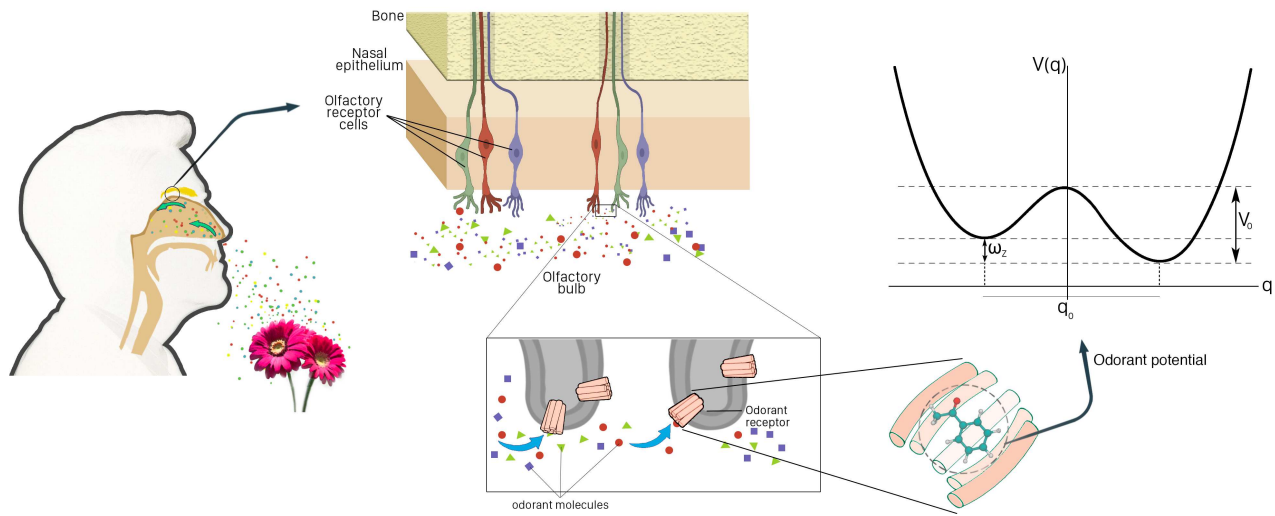}\centering
\caption{A scheme for the sense of smell in which odorants are absorbed by odorant receptors (ORs) in the olfactory receptor cells in the nasal cavity. In the quantum model, each odorant can be simulated as an asymmetric double-well potential for odorant recognition. The signal transduction relies on the success of an electron tunneling from a donor site of an OR to an acceptor site of the same or another OR, facilitated by a vibrational transition in the odorant according to the energy difference between the donor and acceptor sites. }
\end{figure}

Motivated by the phenomena of inelastic electron tunneling (ET) in metals~\cite{Lam,Adk}, Turin proposed that the mechanism of olfactory detection is an odorant-mediated biological inelastic ET~\cite{Tur2}. The signal transduction relies on the success of an ET from a donor (D) site of an OR to an acceptor (A) site of the same or another OR, facilitated by a vibrational transition in the odorant corresponding to the energy difference between D and A sites. Recently, Brookes and co-workers formulated the idea in a semi-classical model to show that such a mechanism fits the observed features of smell~\cite{Bro}. They found that the rate of odorant-mediated inelastic ET and the elastic ET can be drastically different and surprisingly the inelastic ET is the dominant process for certain parameters of the model. The evidence supporting the vibration-based mechanism was obtained from sophisticated quantum chemistry calculations~\cite{Sol,Bit}. Ch\c{e}ci\'{n}ska and co-workers examined the dissipative role of the environment dynamically in vibration-based model and showed that the strong coupling to the environment can enhance the frequency resolution of the olfactory system~\cite{Che}. The main evidence against this theory is given to be the differentiable smells of chiral odorants: they have identical spectra in an achiral solution, but they have different smells~\cite{Ben}. Recently, we addressed this problem using the master equation approach and showed that the chiral recognition in olfaction might rely on the detection of the chiral interactions between the chiral odorant and ORs~\cite{Tir}. A major prediction of the vibration-based theory is the isotope effect: i.e. isotopes should smell differently. Recent behavioural experiments have revealed that fruit flies~\cite{Fra,Bitt}, honeybees~\cite{Gro} and humans~\cite{Haf,Gan} can distinguish isotopes. Yet, experimental evidence against isotopic discrimination keeps the debate open~\cite{Kel,Blo}.\\
\indent In this paper, we examine the physical plausibility of the odorant-mediated inelastic ET model of olfaction. We focus on a typical vibrational degree of freedom of odorant molecule, known as contorsional vibration, in which an atom or a group of atoms oscillates between the two wells of the potential energy surface~\cite{Tow}. Such a vibration is typically modeled by the motion of a particle in a double-well potential. The biological environment is conveniently represented as a collection of harmonic oscillators. We examine the dynamics of the odorant by using the time-dependent perturbation theory and thereby obtain the corresponding elastic and inelastic ET rates for all possible transitions of the odorant. To test the physical limitations of the model, we analyze the rates in different limits of molecular and environmental variables. For simplicity we set $\hbar=1$ throughout the paper.

\section{Model}
We focus on three parts of the system as the main components of the olfaction model: (1) the {\it odorant}, (2) the {\it electron} which tunnels through the odorant, and (3) the surrounding {\it environment}. In the original vibrational model of olfaction, the relevant vibrational mode is represented by a simple harmonic oscillator~\cite{Bro}. Here, we consider a more realistic vibrational mode of non-planer odorant, known as {\it contorsional} mode, in which an atom or a group of atoms oscillates between the left and right wells of a double-well potential. Unlike the harmonic mode the contorsional mode can be used to charachterize the olfactory chiral reconition~\cite{Tir}. Thus, we model the odorant as an asymmetric double-well potential (see Figure 1). The minima of the potential correspond to the left- and right-handed states, $|L\rangle$ and $|R\rangle$, of the odorant (see Figure 2). The handed states can be inter-converted by the quantum tunneling through the barrier $V_{0}$. In the limit $V_{0}\gg \omega_{0}\gg k_{B}T$ ($\omega_{0}$ is the vibration frequency at the bottom of each well), the state space of the odorant is effectively confined in a two-dimensional Hilbert space spanned by two handed states. Such an approximation works properly for a large class of odorants even in the high-temperature limit~\cite{Tow,Her}. The odorant's Hamiltonian can then be expanded by the handed states as $\hat{H}_{od}=-\omega_{z}\hat{\sigma}_{z}+\omega_{x}\hat{\sigma}_{x}$ where $\hat{\sigma}_{i}$ is the $i$-component of Pauli operator and $\omega_{x}$ and $\omega_{z}$ are the tunneling and asymmetry frequencies, respectively. The tunneling frequency $\omega_{x}$ can be calculated from the WKB method as $\omega_{x}=Aq_{0}\sqrt{M\omega_{0}}\exp(-BV_{0}/\omega_{0})$~\cite{Leg}, where $M$ is the molecular mass and $q_{0}$ is the distance between two minima of the potential. The value of the parameters $A$ and $B$ depends on the explicit mathematical form of the potential, but it can usually be approximated by $1$~\cite{Leg,Wei}. The asymmetry is due to the fundamental parity-violating interactions~\cite{Qua,Lee} and the chiral interactions (i.e. interactions that are transformed as pseudoscalars~\cite{Bar}) between the odorant and environmental molecules. The former is typically small but the latter can be significant especially between a chiral odorant and ORs. The eigenstates of the odorant's Hamiltonian can be written as the superposition of handed states as $|E_{1}\rangle=\sin\theta|L\rangle+\cos\theta|R\rangle$ and $|E_{2}\rangle=\cos\theta|L\rangle-\sin\theta|R\rangle$ where we defined $\theta=(1/2)\arctan(\omega_{x}/\omega_{z})$.\\
\indent The electron tunnels through the odorant from a donor state $|D\rangle$ with energy $\varepsilon_{D}$ to an acceptor state $\varepsilon_{A}$ with energy $\varepsilon_{A}$. We then describe the electron with Hamiltonian $\hat{H}_{e}=\varepsilon_{A}|A\rangle\langle A|+\varepsilon_{D}|D\rangle\langle D|$. The biological environment is typically modeled as a collection of harmonic oscillators with Hamiltonian
$\hat{H}_{env}=\sum_{i}\omega_{i}\hat b_{i}^{\dag}\hat b_{i}$ where $b_{i}^{\dag}$ and $b_{i}$ are the creation and annihilation operators for modes of frequency $\omega_{i}$ in the environment.\\
\indent The interaction Hamiltonian has three contributions: between donor and acceptor of the receptor with tunneling strength $\Delta$, between the donor (acceptor) and the odorant with coupling frequency $\gamma_{D}$ ($\gamma_{A}$), and between the donor (acceptor) sites and $i$-the harmonic oscillator of the environment with coupling frequency $\gamma_{iD}$ ($\gamma_{iA}$). Thus, the interaction Hamiltonian of the whole system is given by~\cite{Bro}
\begin{equation}\label{int}
\hat{H}_{int}=\Delta(|A\rangle\langle A|+|D\rangle\langle D|)+(\gamma_{D}|D\rangle\langle D| +\gamma_{A}|A\rangle\langle A|)\otimes\hat{\sigma}_{x}+\sum_{i}(\gamma_{i,D}|D\rangle\langle D|+\gamma_{i,A}|A\rangle\langle A|)\otimes(\hat b_{i}^{\dag}+\hat b_{i})
\end{equation}
The total Hamiltonian characterizes the time evolution of whole system by which the ET rates are calculated. The details of the dynamics are explained in the Methods section.
\section{Results}
\noindent {\bf Tunneling Rates} When the electron tunnels through the odorant, three type of transitions might take place in the odorant (FIG.2): $|L\rangle\rightarrow|E_{2}\rangle$, $|R\rangle\rightarrow|E_{2}\rangle$ and $|E_{1}\rangle\longrightarrow|E_{2}\rangle$.

\begin{figure}[H]
\includegraphics[scale=0.5]{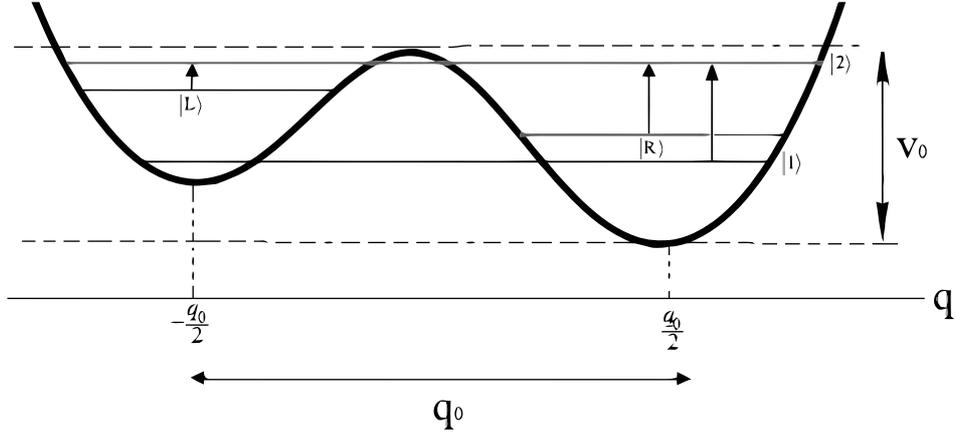}\centering
\caption{Possible transitions of the odorant described as a particle in an asymmetric double-well potential.}
\end{figure}

\noindent The ET rates are obtained from the corresponding probabilities (see Methods) by using $\Gamma_{i\rightarrow j}=dPr_{i\rightarrow j}/dt$. If we set $t_{1}=0$, in the high-temperature limit where the environment is presumably in thermal equilibrium, the inelastic ET rates are given by
\begin{align}\label{L2}
\Gamma_{{D,L\rightarrow A,E_{2}}}=\Delta^{2}\sqrt{\frac{\pi}{k_{B}TJ_{0}\lambda}}\cos(\theta+\upsilon)&\Big[\cos\theta\cos\upsilon  \exp\Big\{\frac{-(\epsilon-J_{0}\lambda+(\eta_{D}-\eta_{A}))^{2}}{4k_{B}TJ_{0}\lambda}\Big\}\nonumber \\
&-\sin\theta\sin\upsilon\exp\Big\{\dfrac{-(\epsilon-J_{0}\lambda-(\eta_{A}+\eta_{D}))}{4k_{B}TJ_{0}\lambda}\Big\}\Big]
\end{align}
\begin{align}\label{R2}
\Gamma_{{D,R\rightarrow A,E_{2}}}=\Delta^{2}\sqrt{\frac{\pi}{k_{B}TJ_{0}\lambda}}\sin(\theta+\upsilon)&\Big[\sin\theta\cos\upsilon  \exp\Big\{\frac{-(\epsilon-J_{0}\lambda-(\eta_{D}-\eta_{A}))^{2}}{4k_{B}TJ_{0}\lambda}\Big\}\nonumber \\
&+\cos\theta\sin\upsilon\exp\Big\{\frac{-(\epsilon-J_{0}\lambda+(\eta_{A}+\eta_{D}))^{2}}{4k_{B}TJ_{0}\lambda}\Big\}\Big]
 \end{align}
 \begin{align}\label{12}
 \Gamma_{{D,E_{1}\rightarrow A,E_{2}}}=\Delta^{2}\sqrt{\frac{\pi}{k_{B}TJ_{0}\lambda}}&\Bigg\{\sin^{2}(\upsilon)\Big[\cos^{2}\theta  \exp\Big\{\frac{-(\epsilon-J_{0}\lambda-(\eta_{A}+\eta_{D}))^{2}}{4k_{B}TJ_{0}\lambda}\Big\}+\sin^{2}\theta
 \exp\Big\{\frac{-(\epsilon-J_{0}\lambda+(\eta_{A}+\eta_{D}))^{2}}{4k_{B}TJ_{0}\lambda}\Big\}\Big]\nonumber\\
 &\!\!\!\!\!\!\!\!\!\!\!\!+\frac{1}{4}\sin\theta\sin(\upsilon)\Big[\exp\Big\{\dfrac{-(\epsilon-J_{0}\lambda-(\eta_{D}-\eta_{A}))^{2}}{4k_{B}TJ_{0}\lambda}\Big\}- \exp\Big\{\frac{-(\epsilon-J_{0}\lambda+(\eta_{D}-\eta_{A}))^{2}}{4k_{B}TJ_{0}\lambda}\Big\}\Big]\Bigg\}
\end{align}
where $\upsilon=\frac{1}{2}[\tan^{-1}(\frac{\omega_{x}+\gamma_{A}}{\omega_{z}})+\tan^{-1}(\frac{\omega_{x}-\gamma_{D}}{\omega_{z}})]$. The elastic ET coincides with the situation where the electron tunnels from the donor site to the acceptor site, without any transition in the odorant. This situation can be considered to be equivalent to the ET in the absence of the odorant~\cite{Che}. The elastic ET rate in the high-temperature limit is given by
\begin{equation}\label{DA}
\Gamma_{{D\rightarrow A}}=\Delta^{2}\sqrt{\dfrac{\pi}{k_{B}TJ_{0}\lambda}} \exp\Big\{\dfrac{-(\epsilon-J_{0}\lambda)^{2}}{4k_{B}TJ_{0}\lambda}\Big\}
\end{equation}
{\bf Physical Parameters} To examine the obtained ET rates quantitatively we first analyze the parameters of the model. Since we aim to examine the model in experiment, the ET rates should be analyzed in terms of controllable parameters (aka variables). These variables include odorant's parameters (tunneling frequency $\omega_{x}$ and asymmetry frequency $\omega_{z}$), and thermodynamical parameters of the environment (temperature and pressure). The parameters with interaction character naturally depend on the odorant's parameters. The energy conservation requires that the energy gap between the donor and acceptor sites $\varepsilon$ be close to the mean value of energy gap between odorant's states. Thus we assume that $\varepsilon\simeq\sqrt{\omega_{x}^{2}+\omega_{z}^{2}}$. The coupling between donor and acceptor sites of OR(s) is weak in comparison with the natural frequency of the odorant, so we estimate $\Delta\simeq0.01\sqrt{\omega_{x}^{2}+\omega_{z}^{2}}Hz$~\cite{Tir}. The coupling frequency between the DA pair and odorant, calculated from the Huang-Rhys factor~\cite{Bro}, is approximated as $\gamma_{D}=-\gamma_{A}\approx0.1\sqrt{\omega_{x}^{2}+\omega_{z}^{2}} $~\citep{Tir}. Since the biological environment is microscopically uncontrollable, the parameters of the corresponding spectral density is considered as mere parameters. The most common biological environment is water. The parameters of an aqueous environment can be estimated as $J_{0}\approx1$ and $\lambda\approx10^{12}Hz$~\cite{Gil,Tir2}.\\
{\bf Odorant Analysis} We examine the ET rates for different odorants in terms of their molecular parameters, $\omega_{x}$ and $\omega_{z}$. The magnitude of tunneling frequency $\omega_{x}$, ranging from the inverse of the lifetime of the universe to millions of hertz, can be extracted from the spectroscopic data~\cite{QuSt}. The asymmetry frequency of the odorant, $\omega_{z}$, represents an overall measure of all chiral interactions involved. For our system, these interactions are primarily due to the intermolecular interactions between the odorant and ORs. The magnitude of intermolecular interactions can in principle be determined by using quantum chemistry calculations. The dependence of the ET rates to $\omega_{x}$ and $\omega_{z}$ for different transitions of the odorant is plotted in Figure 3. This clearly shows that the vibrational model based on odorant-mediated inelastic ET is improbable for a wide range of odorants.

\begin{figure}[H]
  \subfigure[]{\includegraphics[scale=0.6]{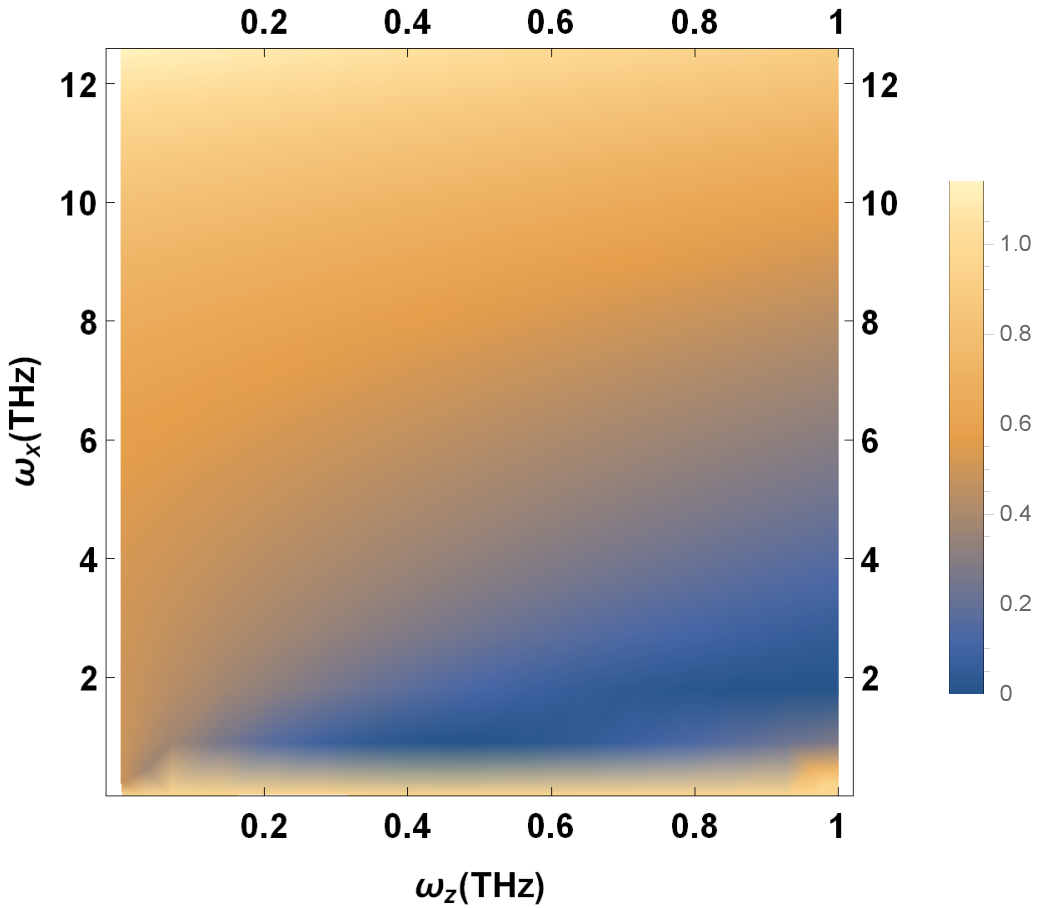}}\centering
  \subfigure[]{\includegraphics[scale=0.6]{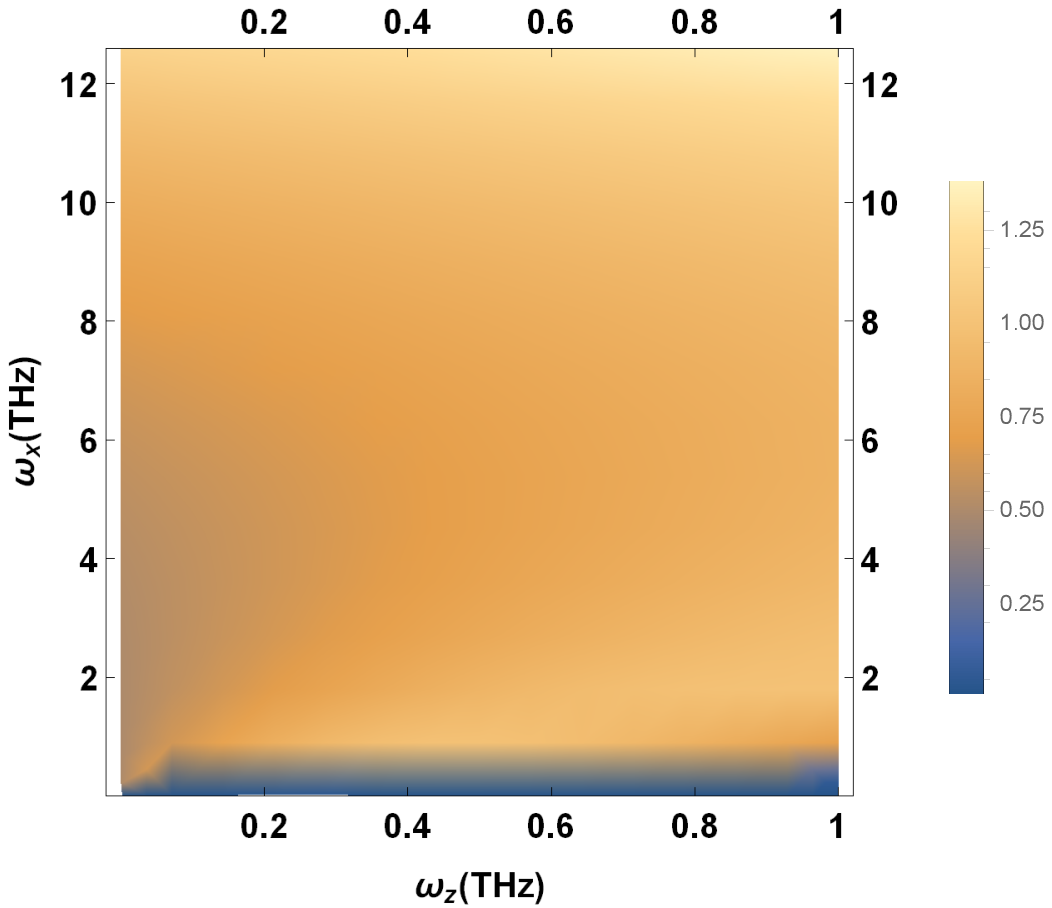}}\centering
  \subfigure[]{\includegraphics[scale=0.6]{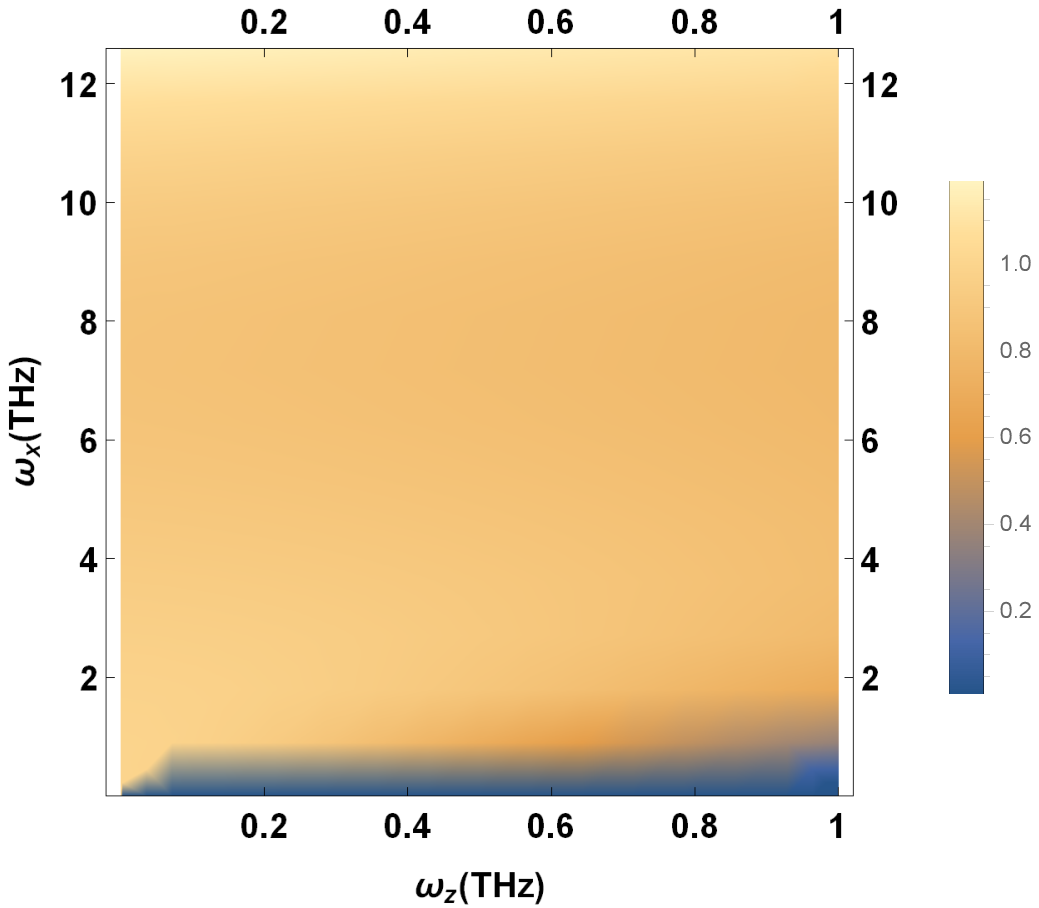}}\centering
  \caption{The inelastic-to-elastic ratio versus tunneling frequency $\omega_{x}$ and asymmetry frequency $\omega_{x}$ at biological temperature $T=310K$ for transitions (a) $L\rightarrow E_{2}$, (b) $R\rightarrow E_{2}$ , and (c) $E_{1}\rightarrow E_{2}$.}
\end{figure}

\noindent Since ORs are chiral structures, we assume that the contribution of chiral interactions is significant. At a fixed high magnitude of asymmetry parameter $\omega_{z}$, the inelastic-to-elastic ratio against the tunneling frequency $\omega_{x}$ are plotted in Figure 4 for different transitions of the odorant. Two different behaviors can be identified here; in the asymmetry-dominant limit, $\omega_{x}<\omega_{z}$, although different transitions exhibit different dependencies on $\omega_{x}$, the inelastic ET is not dominant. In the tunneling-dominant limit, $\omega_{x}\geq\omega_{z}$, however, for all transitions the inelastic ET is dominant. In other words, for each transition there is a threshold of tunneling frequency $\omega_{x}$ in the bottom limit in which the olfactory system cannot recognize the odorant. This fact can be used to examine the model in experiment.

\begin{figure}
\includegraphics[scale=0.4]{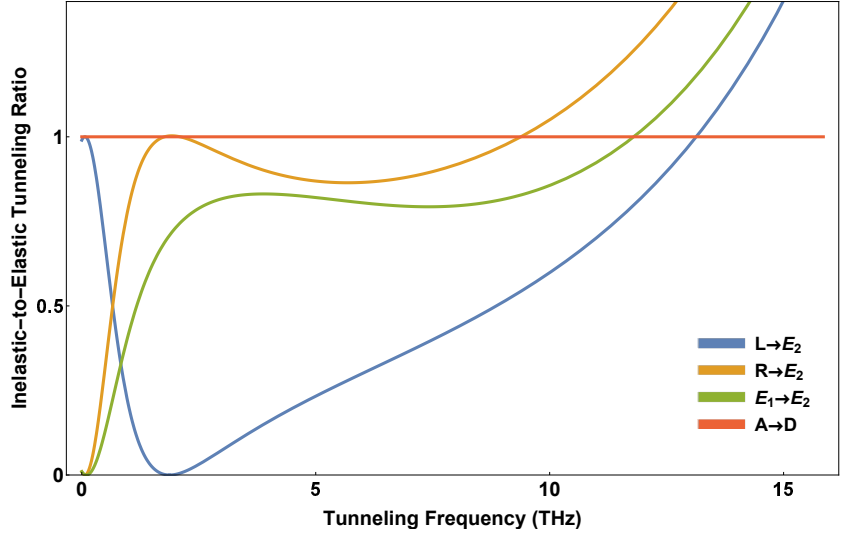}\centering
\caption{The inelastic-to-elastic ratio versus the tunneling frequency $\omega_{x}$ at biological temperature $T=310K$ for transitions $L\rightarrow E_{2}$, $R\rightarrow E_{2}$ and $E_{1}\rightarrow E_{2}$. All figures are plotted at $\omega_{z}=10^{12}Hz$.}
\end{figure}

\noindent {\bf Temperature Analysis} The temperature dependency of different transitions of the odorant are essentially similar. At a fixed high magnitude of asymmetry parameter $\omega_{z}$, for the transitions $L\rightarrow E_{2}$, $R\rightarrow E_{2}$, and $E_{1}\rightarrow E_{2}$ the inelastic-to-elastic ratio versus the tunneling frequency $\omega_{x}$ are plotted in Figure 5 for different temperatures of the environment. For each odorant (with a fixed tunneling frequency $\omega_{x}$) there is a threshold for temperature in the bottom limit in which the olfactory system cannot recognize the scent. This fact can also be used to examine the model in experiment.

\begin{figure}[H]
\subfigure[]{\includegraphics[scale=0.22]{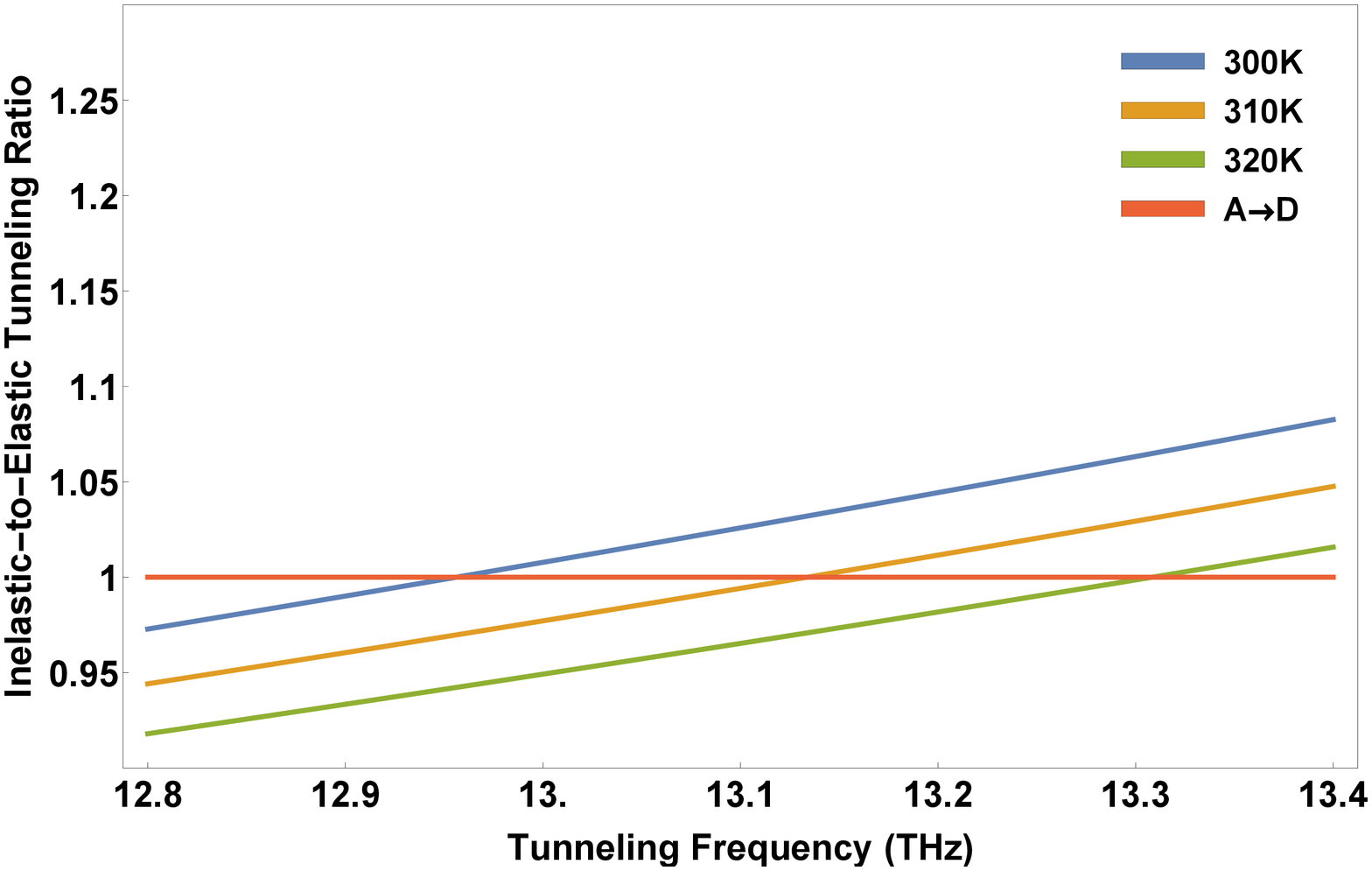}}\centering
\subfigure[]{\includegraphics[scale=0.22]{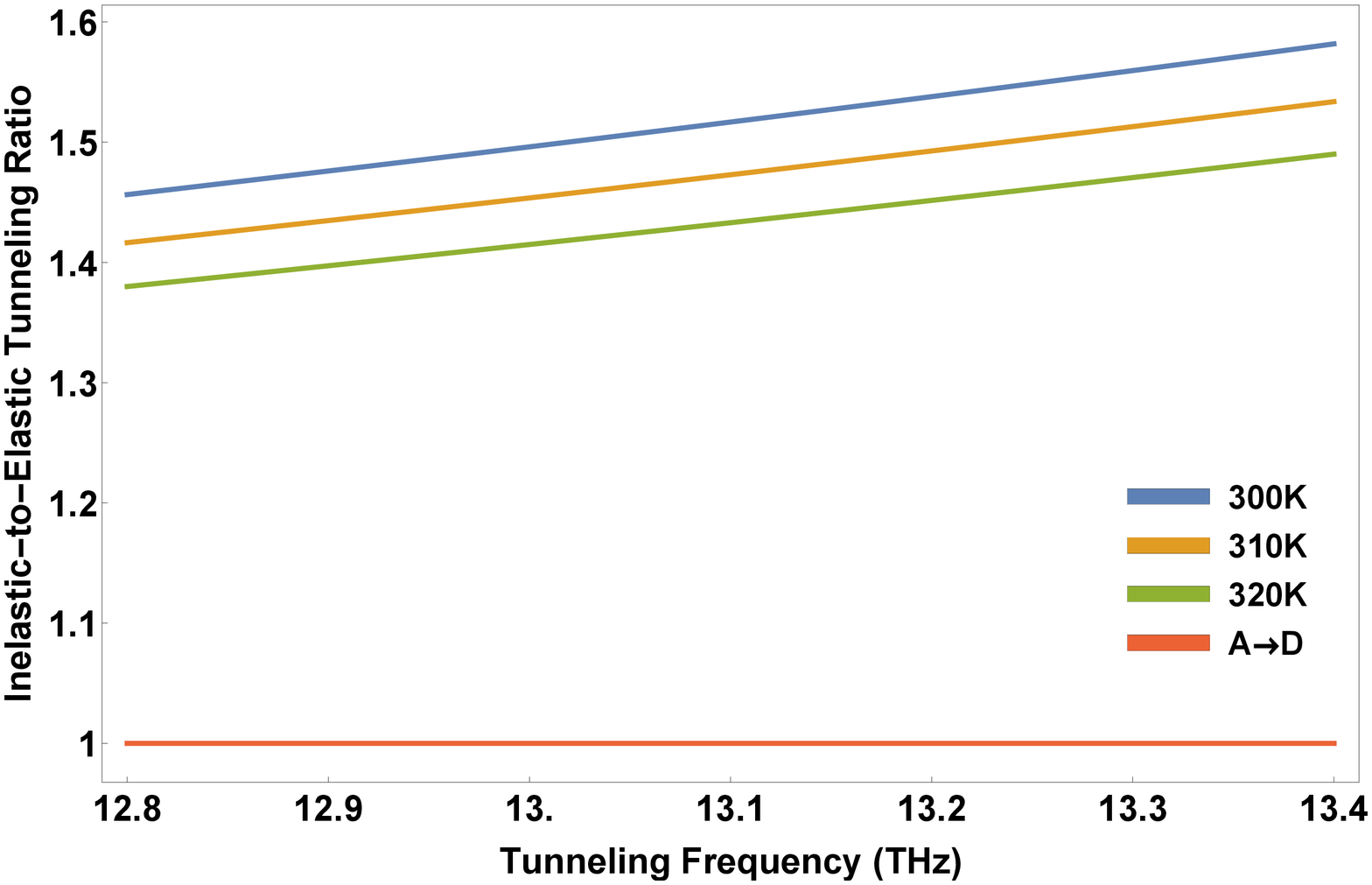}}\centering
\subfigure[]{\includegraphics[scale=0.22]{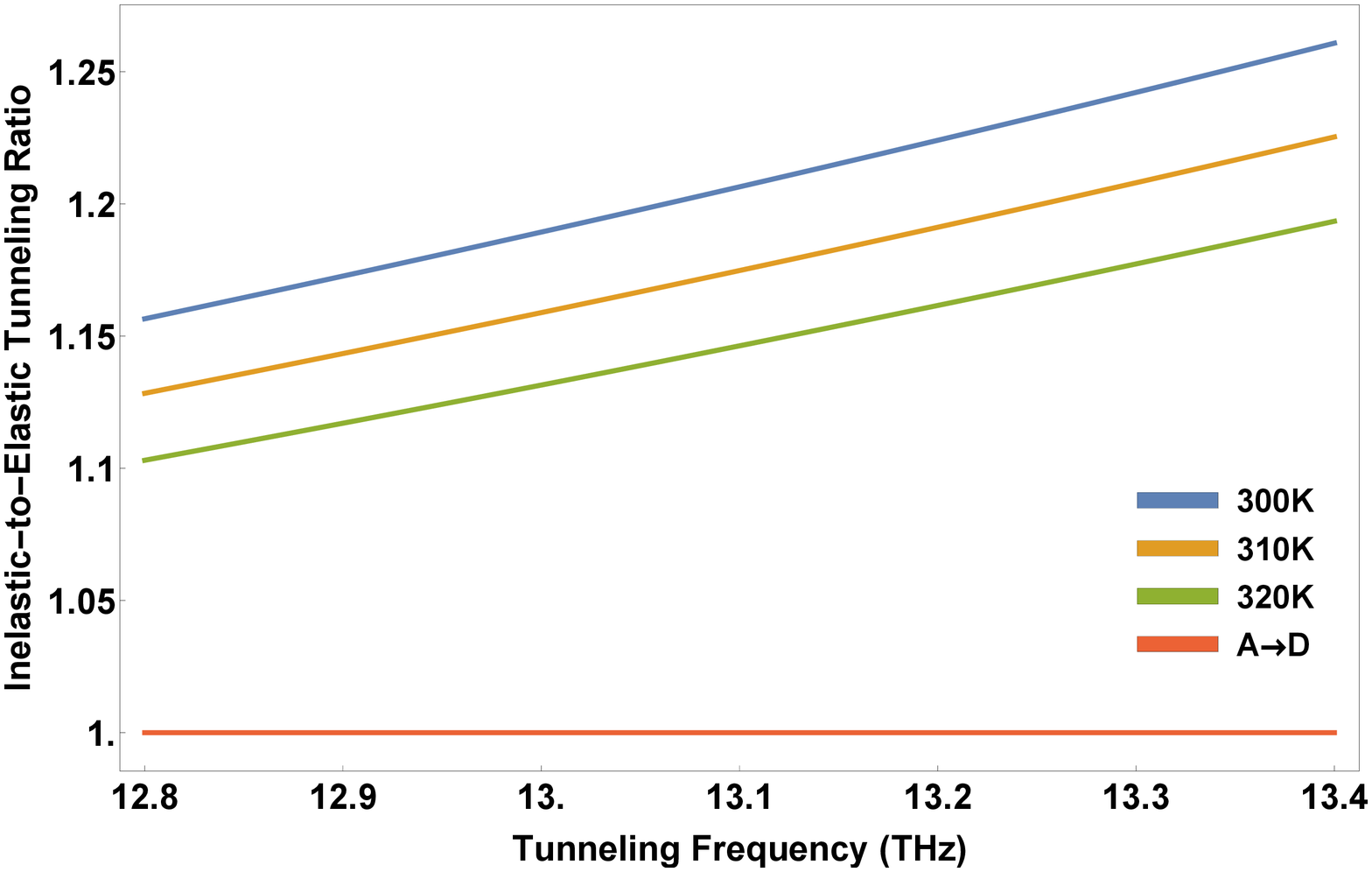}}\centering
\caption{The inelastic-to-elastic ratio versus the tunneling frequency $\omega_{x}$ at temperatures $T=300$K, $310$K, and $320$K for transitions (a) $L\rightarrow E_{2}$, (b) $R\rightarrow E_{2}$, and (c) $E_{1}\rightarrow E_{2}$. All figures are plotted at $\omega_{z}=10^{12}$Hz.}
\end{figure}

\noindent {\bf Pressure Analysis} The concentration of the odorant in the condensed environment is proportional to its pressure. The tunneling frequency of odorants is related to the pressure of the environment. To illustrate the pressure dependency of the odorant's dynamics, we focus on the ammonia molecule $NH_{3}$ as odorant. In the low-pressure limit, the tunneling frequency of ammonia, known as inversion frequency, is estimated as $\omega_{x}\simeq 2.4\times10^{10}Hz$~\cite{Ble}, and at $P\simeq 2atm$, $\omega_{x}$ shifts to zero. This phenomenon is theoretically demonstrated in the context of the mean-field theory~\cite{Jon}. The pressure dependency of the inversion frequency is given by $\omega_{x}^{\prime}=\omega_{x}\sqrt{1-P/P_{cr}}$ where the critical pressure $P_{cr}$ is approximately $1.6~atm$ at room temperature. If $P\rightarrow P_{cr}$, then $\omega^{\prime}\rightarrow0$, and accordingly the handed states become eigenstates of the molecular Hamiltonian. In the high-pressure limit, the relevant transition in the odorant would be $|R\rangle\rightarrow|L\rangle$ (see Figure 2). The inelastic ET rate according to this transition is obtained as
\begin{equation}\label{LR}
\Gamma_{{D,R\rightarrow A,L}}=\Delta^{2}\sqrt{\frac{\pi}{k_{B}TJ_{0}\lambda}}\sin^{2}\upsilon\exp\Big\{ \frac{-(\epsilon-J_{0}\lambda+(\eta_{1}+\eta_{2}))^{2}}{4k_{B}TJ_{0}\lambda}\Big\}
\end{equation}
The inelastic-to elastic ratio versus the tunneling frequency $\omega_{x}$ in the high-pressure limit is plotted in Figure 6 for $\omega_{z}=1$ THz, 5 THz, and 10 THz respectively.

\begin{figure}[H]
\subfigure[]{\includegraphics[scale=0.215]{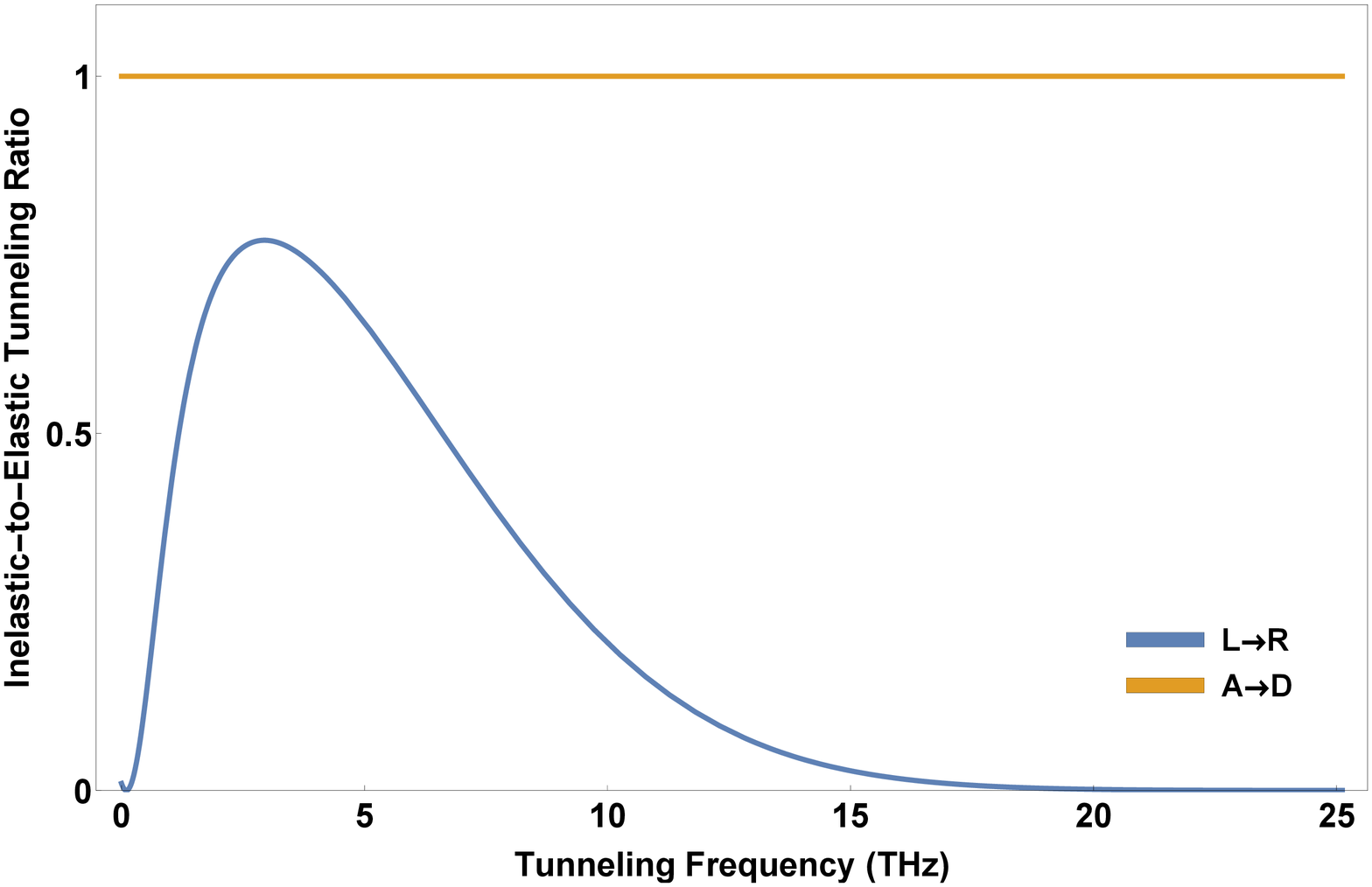}}\centering
\subfigure[]{\includegraphics[scale=0.21]{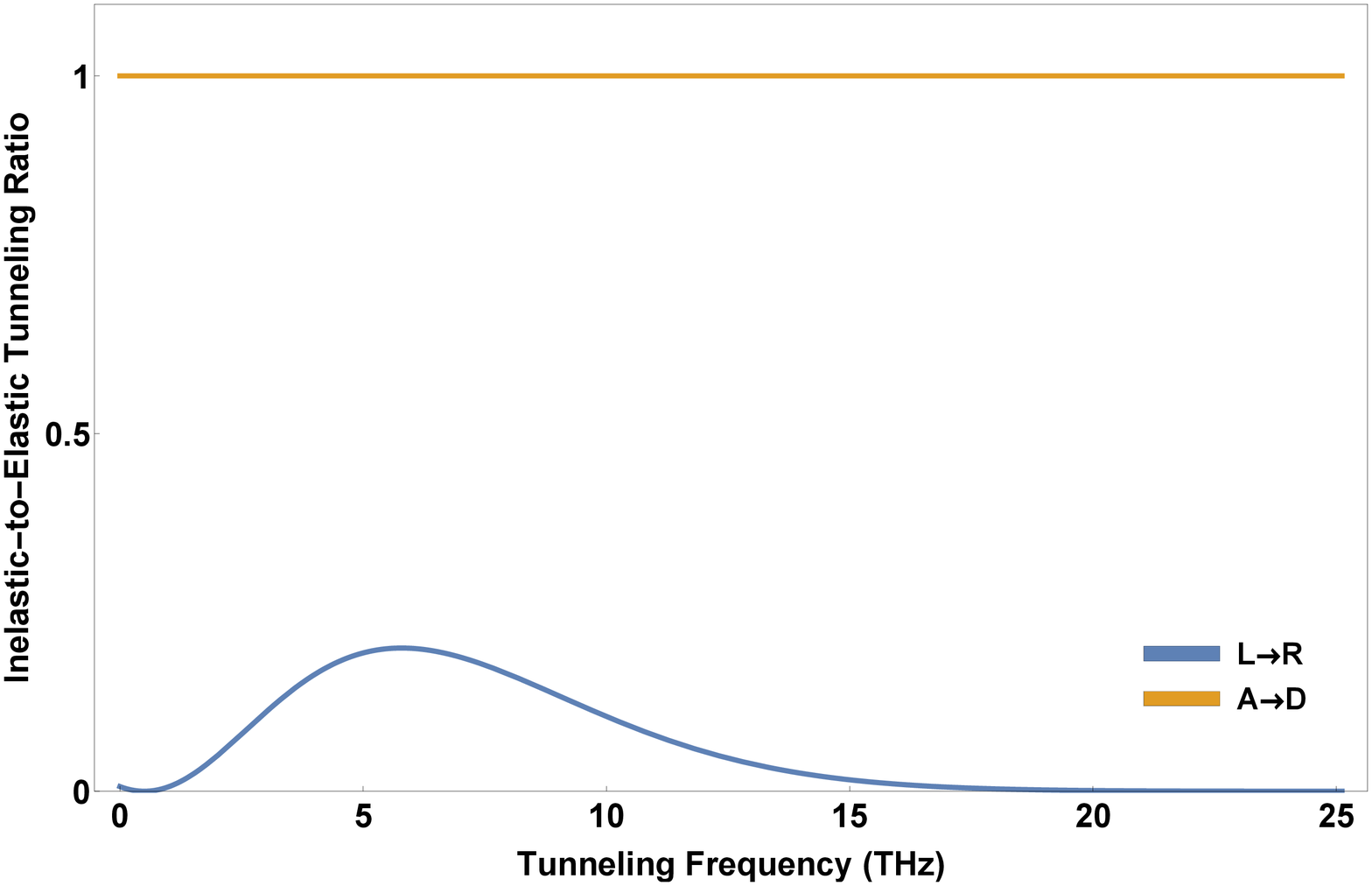}}\centering
\subfigure[]{\includegraphics[scale=0.22]{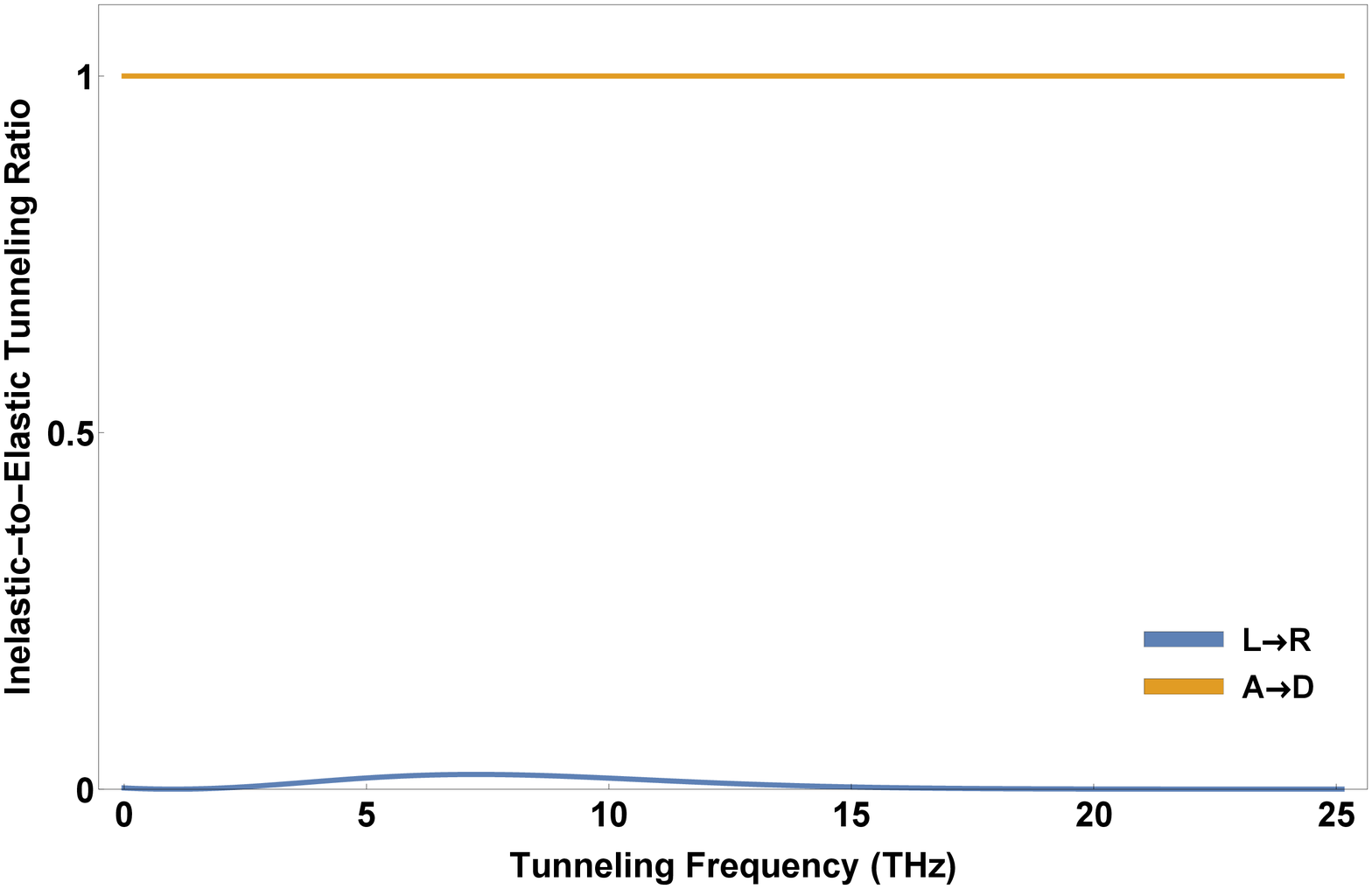}}\centering
\caption{The inelastic-to-elastic ratio versus the tunneling frequency $\omega_{x}$ in the high-pressure limit for transition $DR\rightarrow AL$ for (a) $\omega_{z}=1$ THz, (b) $\omega_{z}=5$ THz, and (c) $\omega_{z}=10$ THz.}
\end{figure}

\noindent In the high-pressure limit, the inelastic ET is always ineffective (see Figure 6). This provides another empirical test for probing the validity of the tunneling model of olfaction. Specially, we predict that at $P\geq P_{cr}$ the olfactory system is unable to recognize the smell if the tunneling mechanism is at work.\\

{\bf Isotopic Effect} The isotopic dependency of ET rates can be followed from the mathematical form of critical pressure as $P_{cr}=\sqrt{2\omega_{x}^{2}mk_{B}T}/\sigma(P)$~\cite{Wea} where we defined $m=\mu m_{t}/(\mu+m_{t})$ with $\mu$ as the reduced mass of the odorant, $m_{t}$ as a measure of average mass of the odorant and environment, and $\sigma(P)$ is the decoherence cross-section. Using the hard sphere model for the odorant at room temperature, the decoherence cross-section can be approximated as 150$a_{b}$ ($a_{b}$ is the Bohr radius). Isotopic substitution alters $m$ and its effect is significant for heavier odorants. Replacing an isotope with another with larger mass causes increasing $P_{cr}$. According to $\omega_{x}^{\prime}=\omega_{x}\sqrt{1-P/P_{cr}}$ if $P_{cr}$ is increased then $\omega_{x}^{\prime}$ approaches to $\omega_{x}$ and we arrive the limit in which the inelastic ET is effective. As a result, the substitution of massive isotopes amplifies the sensitivity of possible quantum mechanism of odorant discrimination. For large odorants, the sensitivity is affected considerably. Exact values of $P_{cr}$ can be calculated for different odorants using computational methods. So, we expect that the ability to distinguish a specific odorant is decreased when exposed to the pressures near $P_{cr}$. Moreover, when we replace atoms of the odorant with their heavier isotopes, we predict that the same behavior appears in pressures larger than the $P_{cr}$ of the original odorant.\\

{\bf Chiral Recognition} The existence of enantiomers, which have different smells, are usually used as an argument to reject the quantum model of olfaction~\cite{Bit}. Since such enantiomers have the same vibrational spectrum, it seems that the shape-based parameters should be included in the quantum model to distinguish them from each other. The model presented here can be used to generalize the vibrational model for chiral recognition. Chiral molecules can be effectively modeled by a double-well potential~\cite{Her}. The shape factor is the configuration of the chiral odorant. Our model predicts that the ET rates of inelastic ET are different for the two enantiomers for transitions $|L\rangle\rightarrow|E_{2}\rangle$ and $|R\rangle\rightarrow|E_{2}\rangle$ (see equations (\ref{L2}) and (\ref{R2})). Our results here are in agreement with the Born-Markov master equation approach~\cite{Tir}. The values of elastic and inelastic ET rates are given in TABLE 1 for a series of odorant parameters. The enantiomers with similar smells lie in the limit $\omega_{x}\ll\omega_{z}$ (i.e. the first block in the TABLE 1). But the inelastic ET is ineffective in this limit. The enantiomers have different smells in the limit $\omega_{x}\approx\omega_{z}$ (i.e. the second block in the TABLE 1), but inelastic ET is still ineffective. The inelastic ET is effective for all transitions in the limit $\omega_{x}>\omega_{z}=\lambda$ (i.e. the third row of third block in the TABLE 1). In this limit, the ratio of the inelastic ET rate for the left-handed enantiomer to that of the right-handed one increases with the ratio of the tunneling frequency to the asymmetry frequency. Typical times for electron transfer in proteins are of order $10^{-15}-10^{-12}s$~\cite{Bre}. Although the difference between inelastic ET rates of transitions may be appeared insignificant, however, in comparison with similar processes in biology it can be possible for the system to discriminate between two enantiomers under quantum constraints.

\begin{table}[H]
  \begin{center}
  \def\arraystretch{1.2}
  \caption{Elastic and inelastic ET rates for some parameters of the chiral odorants at biological temperature $T=310K$.}
  \begin{tabular} {|c|c|c|c|c|}
  \hline
   $\omega_{x}$ &\hspace{1cm}$\omega_{z}$ &\hspace{1cm} $\Gamma_{D\rightarrow A}^{-1}$ &\hspace{1cm} $\Gamma_{D,L\rightarrow A,E_{2}}^{-1}$ &\hspace{1cm} $\Gamma_{D,R\rightarrow A,E_{2}}^{-1}$ \\
  Hz &\hspace{1cm} Hz &\hspace{1cm} s &\hspace{1cm} s &\hspace{1cm} s \\
  \hline
  \hline
  $10^{3}$ &\hspace{1cm} $10^{6}$ &\hspace{1cm} 15049.3 &\hspace{1cm} 15195.3 &\hspace{1cm} $1.56632\times10^{6}$ \\
  \hline
  $10^{6}$ &\hspace{1cm} $10^{9}$ &\hspace{1cm} 0.0150482 &\hspace{1cm} 0.0151942 &\hspace{1cm} 1.56632 \\
  \hline
  $10^{9}$ &\hspace{1cm} $10^{12}$ &\hspace{1cm} $1.44896\times10^{-8}$ &\hspace{1cm} $1.46302\times10^{-8}$ &\hspace{1cm} $1.56721\times10^{-6}$ \\
  \hline
  $10^{12}$ &\hspace{1cm} $10^{13}$ &\hspace{1cm} $3.19822\times10^{-9}$ &\hspace{1cm} $3.20438\times10^{-9}$ &\hspace{1cm} $1.65938\times10^{-6}$ \\
  \hline
  \hline
  $10^{3}$ &\hspace{1cm} $10^{3}$ &\hspace{1cm} $7.52468\times10^{9}$ &\hspace{1cm} $3.68835\times10^{10}$ &\hspace{1cm} $9.45326\times10^9$ \\
  \hline
  $10^{6}$ &\hspace{1cm} $10^{6}$ &\hspace{1cm} 7524.68 &\hspace{1cm} 36883.5 &\hspace{1cm} 9453.26 \\
  \hline
  $10^{9}$ &\hspace{1cm} $10^{9}$ &\hspace{1cm} 0.00752388 &\hspace{1cm} 0.0368776 &\hspace{1cm} 0.00945288 \\
  \hline
$10^{12}$ &\hspace{1cm} $10^{12}$ &\hspace{1cm} $7.29208\times10^{-9}$ &\hspace{1cm} $3.5273\times10^{-8}$ &\hspace{1cm} $9.31282\times10^{-9}$ \\
  \hline
  \hline
  $10^{6}$ &\hspace{1cm} $10^{3}$ &\hspace{1cm} 15049.3 &\hspace{1cm} 30196. &\hspace{1cm} 30002. \\
  \hline
  $10^{9}$ &\hspace{1cm} $10^{6}$ &\hspace{1cm} 0.0150482 &\hspace{1cm} 0.0151942 &\hspace{1cm} 1.56632 \\
  \hline
  $10^{12}$ &\hspace{1cm} $10^{9}$ &\hspace{1cm}$1.44896\times10^{-8}$ &\hspace{1cm} $2.99804\times10^{-8}$ &\hspace{1cm} $2.97857\times10^{-8}$ \\
  \hline
$10^{13}$ &\hspace{1cm} $10^{12}$ &\hspace{1cm}$3.19822\times10^{-9}$ &\hspace{1cm} $3.69778\times10^{-10}$ &\hspace{1cm} $2.38954\times10^{-10}$ \\
\hline
  \end{tabular}
 \hfill
  \end{center}
\end{table}

\section{Summary}
Our analysis can be summarized as follows for the main ingredients of the olfactory system: \\
\begin{itemize}
  \item {\it Odorant}: In the original vibrational model of olfaction, the relevant vibrational mode is represented by a simple harmonic oscillator~\cite{Bro}. Here, we focused on a more realistic vibrational mode of non-planer odorant, known as {\it contorsional} mode, in which an atom or a group of atoms oscillates between the left and right wells of a double-well potential. Unlike the harmonic mode the contorsional mode can be used to charachterize the olfactory chiral reconition~\cite{Tir}. The eigenstates of the double-well potential are essentially doplets. The first doplet is energetically available for most molecules at room temperature~\cite{Tow,Her}. The two-dimensional Hamiltonian of the mode is expressed by the tunneling frequency $\omega_{x}$ and asymmetry frequency $\omega_{z}$. Our expressions for the (in-)elastic ET rates are reduced to the corresponding expressions for a harmonic mode in the limit $\omega_{x}\rightarrow0$~\cite{Che}.
\item {\it Electron}: The detailed biological origin of the electron which tunnels through the odorant is not known but it may be due to redox agents in the cell fluid~\cite{Row}. According to the original model~\cite{Bro}, we considered donor (D) and acceptor (A) sites of traveling electron as single molecular orbitals with energies $\varepsilon_{D}$ and $\varepsilon_{D}$, coupled to each other by a weak hopping integral $\Delta$. In order to satisfy energy conservation during the tunneling process, the electron's parameters should be consistent with the odorant's paramaters. Thus, they cannot be considered as variables.
  \item {\it Environment}: We modeled the biological environment as a harmonic bath with an ohmic spectral density according to the original model~\cite{Bro}. Such an environment is characterized by its microscopic parameters (e.g. coupling frequency $ J_{0}$ and cut-off frequency $\lambda$) and macroscopic parameters (e.g. temperature and pressure). Unlike the microscopic parameters, the macroscopic parameters can be controlled in experiment and thus they are considered as variables.
\end{itemize}

\section{Conclusion}
In this paper, we have generalized and proposed an analysis for examination of the dissipative quantum model of olfaction in experiments. In fact, it has been suggested that inelastic electron tunneling through the odorant potential is a dominant process in olfaction. Here, we have suggested a region of easy measurable parameters in the lab (e.g. temperature and pressure) in which we can discriminate between the elastic and inelastic tunneling through the potential. We have shown that the dissipative odorant-mediated inelastic electron tunneling mechanism of olfaction is ineffective for a wide range of odorants, and the range of ineffectiveness depends on the type of transition in the contortional mode. Moreover, our results indicate that there are thresholds in the bottom limit for both temperature and pressure in which the olfactory system cannot recognize the scent. Additionally, the substitution of massive isotopes amplifies the sensitivity of olfactory odorant discrimination. Perhaps the most relevant part of our analysis is related to the chiral recognition of the odorants in which enantiomers with similar smells lie in the asymmetry-dominant limit of dynamics and enantiomers with different smells lie in the tunneling-dominant limit of dynamics.\\
\indent We expect that our analysis can be used in experiments to examine the validity of the dissipative quantum model of olfaction.

\section{Methods}
\noindent {\bf Polaron Transformation} The unperturbed Hamiltonian $\hat H_{0}=\hat{H}_{od}+\hat{H}_{e}+\hat{H}_{env}$ is diagonalized by a polaron transformation as~\cite{Tir}
\begin{equation}\label{MI1}
\hat H'_{0}=\sum_{R=A,D}(\varepsilon_{R}+\eta_{R}\hat{\sigma}_{z})|R\rangle\langle R|+\sum_{i}\omega_{i}\hat b_{i}^{\dag}\hat b_{i}
\end{equation}
where
\begin{align}\label{MI2}
 \eta_{A}&=-\frac{\omega_{z}}{2}\cos\big\{\tan^{-1}\big(\dfrac{\omega_{x}+\gamma_{A}}{\omega_{z}}\big)\big\}\nonumber\\
 \eta_{D}&=-\frac{\omega_{z}}{2}\cos\big\{\tan^{-1}\big(\dfrac{\omega_{x}-\gamma_{D}}{\omega_{z}}\big)\big\}
\end{align}
Similarly, the interaction Hamiltonian is transformed to
\begin{equation}\label{MI3}
\hat H'_{int}=\Delta |A\rangle\langle D|\exp\Big\{\dfrac{i}{2}\big[\tan^{-1}\big(\dfrac{\omega_{x}+\gamma_{A}}{\omega_{z}}\big)+
\tan^{-1}\big(\dfrac{\omega_{x}-\gamma_{D}}{\omega_{z}}\big)\big]\hat\sigma_{y}\Big\})\exp\Big\{\sum_{i}\big(\dfrac{\gamma_{i,D}-\gamma_{i,A}}{\omega_{i}}\big)(\hat b_{i}^{\dag}-\hat b_{i})\Big\}+ h.c
\end{equation}
{\bf Dynamics} Time evolution operator in the interaction picture at weak-coupling limit can be written as
\begin{equation}\label{MII1}
\hat{U}_{I}(t)=1-i \int_{0}^{t}dt_{1}\hat{H}_{int}(t_{1})-\int_{0}^{t}dt^{\prime}_{1}\int_{0}^{t^{\prime}_{1}}dt_{1}\hat{H}_{int}(t'_{1})\hat{H}_{int}(t_{1})
\end{equation}
We assume that initially the electron is located at the donor site $|D\rangle$, the odorant is found in the ground state of energy or left- or right-handed states (see FIG.1), all denoted by $|i\rangle$, and the environment is described by the density matrix $\rho_{env}(0)$. The initial state of the whole system is then $\rho(0)=|D,i\rangle\langle D,i|\rho_{env}(0)$. The density matrix of the whole system at time $t$ is given by $\rho_{I}(t)_=\hat{U}_{I}(t)\rho(0)\hat{U_{I}^{\dag}}(t)$. The probability of finding the electron at time $t$ on the acceptor site and the odorant in the state $|j\rangle$ is obtained as
\begin{align}\label{MII2}
Pr_{D,i\rightarrow A,j}&=Tr_{env}\langle A,j| \hat{U}_{I}(t)\rho(0)\hat{U_{I}^{\dag}}(t)|A,j\rangle\nonumber\\
&=\Delta^{2}\int_{0}^{t}dt^{\prime}_{1}\int_{0}^{t^{\prime}_{1}}dt_{1} e^{-i\epsilon(t_{1}-t^{\prime}_{1})}\langle j|\hat\Omega{(t_{1})}|i\rangle \langle i|\hat\Omega{(t^{\prime}_{1})}|j\rangle f(\omega, t_{1},t^{\prime}_{1})
\end{align}
where $\hat\Omega(t)$ is a matrix with elements
\begin{align}\label{MII3}
\Omega_{11}(t)&=\Omega_{22}(-t)=\cos\Big\{\dfrac{1}{2}\big[\tan^{-1}\big(\dfrac{\omega_{x}+\gamma_{A}}{\omega_{z}}\big)
+\tan^{-1}\big(\dfrac{\omega_{x}-\gamma_{D}}{\omega_{z}}\big )\big]\Big\} e^{-i t(\eta_{2}-\eta_{1})}\nonumber \\
\Omega_{12}(t)&=-\Omega_{21}(-t)=\sin\Big\{\dfrac{1}{2}\big[\tan^{-1}\big(\dfrac{\omega_{x}+\gamma_{A}}{\omega_{z}}\big)
+\tan^{-1}\big(\dfrac{\omega_{x}-\gamma_{D}}{\omega_{z}}\big )\big]\Big\} e^{-i t(\eta_{1}+\eta_{2})}
\end{align}
and $ f(\omega, t_{1},t^{\prime}_{1}) $ is the correlation function of the environment, defined by
\begin{equation}\label{MII4}
f(\omega, t_{1},t^{\prime}_{1})=Tr_{env}\big\{\hat\Theta(t_{1})\rho_{env}(0)\hat\Theta^{\dag}(t^{\prime}_{1})\big\}
\end{equation}
where $\hat\Theta(t)$ is a displacement operator
\begin{equation}\label{MII5}
\hat\Theta(t)=\exp\Big\{\sum_{i}\dfrac{\gamma_{i,D}-\gamma_{i,A}}{\omega_{i}}\Big(e^{i\omega_{i}t}\hat b_{i}^{\dag}-e^{-i\omega_{i}t}\hat b_{i}\Big)\Big\}
\end{equation}
To calculate the environmental correlation function (\ref{MII4}) we should specify the initial state of the environment. Regarding the environment in the thermal equilibrium, the corresponding correlation function is obtained as
\begin{align}\label{MII6}
f(\omega,t_{1},t^{\prime}_{1})&=Tr_{env}\big\{\hat\Theta(t_{1})\Big(\frac{1}{1+\tilde{n}}\sum_{n=0}^{\infty}\big(\dfrac{\tilde{n}}{1+\tilde{n}}\big)^{n}|n\rangle\langle n|\Big)\hat\Theta^{\dag}(t^{\prime}_{1})\big\}\nonumber \\
&=e^{i Im\zeta(t_{1})\zeta(t^{\prime}_{1})}\frac{1}{1+\tilde{n}}\sum_{n=0}^{\infty}(\dfrac{\tilde{n}}{1+\tilde{n}})^{n}\langle n|e^{\sum_{i}\chi_{i}\hat{b^{\dag}}_{i}-\chi^{\ast}_{i}\hat{b_{i}}}|n\rangle
\end{align}
where $\tilde{n}=1/(e^{\omega/k_{B}T}-1)$, $|n\rangle$ is the number state and $\chi_{i}:=\zeta(t_{1})+\zeta(t^{\prime}_{1})$. To obtain a closed mathematical form for correlation function we use the following relation for spectral density $J(\omega)$ of the environmental particles
\begin{equation}\label{MII7}
J(\omega)=\sum_{i}(\gamma_{i,D}-\gamma_{i,A})^{2}\delta(\omega-\omega_{i})\equiv J_{0}\omega e^{-\frac{\omega}{\lambda}}
\end{equation}
where $J_{0}$ is a measure of the coupling between the system and environment, and $\lambda$ is the cut-off frequency of the environmental particles. Inserting (\ref{MII7}) in (\ref{MII6}), correlation function is summed up as
\begin{equation}\label{MII8}
f(\omega,t_{1},t^{\prime}_{1})=\exp\Big\{-\int_{0}^{\infty}\dfrac{J(\omega)}{\omega^{2}}\Big[1-\cos\big(\omega(t_{1}-t^{\prime}_{1})\big)f(\omega)
-i\sin\big(\omega(t_{1}-t^{\prime}_{1})\big)\Big]\Big\}
\end{equation}
where for the environment in the ground state $ f(\omega)=1$ and for the thermal environment $ f(\omega)=\coth(\omega/2k_{B}T)$.

\section{Acknowledgement}
F.T.G acknowledges the financial support of Iranian National Science Foundation (INSF) for this work. V.S. thanks M. Aslani for illustrating the olfactory system in figure 1.

\section{Author Contributions}
All authors contributed to developing the proposal and writing the manuscript.

\section{Additional Information}
The authors declare no competing financial interests.


\begin{thebibliography}{1}
\bibitem{Row} D. J. Rowe, {\it Chemistry and technology of flavors and fragrances}, Blackwell, Oxford, 2005.
\bibitem{Axe} R. Axel, Scents and Sensibility: A Molecular Logic of Olfactory Perception (Nobel Lecture), {\it Angew. Chem. Int. Ed.} {\bf 44}, 6110, 2005.
\bibitem{Buc} L. B. Buck, Unraveling the Sense of Smell (Nobel Lecture), {\it Angew. Chem. Int. Ed.} {\bf 44}, 6128, 2005.
\bibitem{LeeK} S. H. Lee{\it et al.} Mimicking the human smell sensing mechanism with an artificial nose platform, {\it Biomaterials} {\bf 33}, 1722, 2012.
\bibitem{Fara} R. H. Farahi, A. Passian, L. Tetard and T. Thundat, Critical Issues in Sensor Science To Aid Food and Water Safety, {\it ACS
Nano} {\bf 6}, 4548, 2012.
\bibitem{Zar} M. Zarzo, The sense of smell: molecular basis of odorant recognition, {\it Biol. Rev.} {\bf 82}, 455, 2007.
\bibitem{Amo} J. Amoore, The stereochemical theory of olfaction, {\it Nature} {\bf 199}, 912, 1963.
\bibitem{Mor} K. Mori and G. Shepard, Emerging principles of molecular signal processing by mitral/tufted cells in the olfactory bulb, {\it Semin. Cell. Biol.} {\bf 5}. 65, 1994.
\bibitem{Yos} F. Yoshii, S. Hirono and I. Moriguchi, Relations between the odor of (r) ethyl citronellyl oxalate and its stable conformations, {\it Quant. Struc-Act Rel.} {\bf 13} 144147, 1994.
\bibitem{Ara} R. C. Araneda, A. D. Kini and S. Firestein. The molecular receptive range of an odorant receptor, {\it Nature Neuroscience}, {\bf 3}, 1248, 2000.
\bibitem{Tur} L. Turin and F. Yoshii, Struture-odor relations: a modern perspective, in {\it Handbook of Olfaction and Gustaion}, R. Doty, Marcel Dekker, New York, 2003.
\bibitem{Ben} R. Bentley, The nose as a stereochemist: Enantiomers and odor, {\it Chem. Rev.} {\bf 106}, 4099, 2006.
\bibitem{Bro0} J. C. Brookes, A. P. Horsfield, and A. M. Stoneham, Odour character differences for enantiomers correlate with molecular flexibility, {\it J. R. Soc. Interface} {\bf 6}, 75, 2009.
\bibitem{Dys} G. Dyson, The scientific basis of odour, {\it Chem. Ind.} {\bf 57}, 647, 1938.
\bibitem{Wri} R. Wright, Odor and molecular vibrations: neural coding of olfactory information, {\it J. Theor. Biol.} {\bf 64}, 473, 1977.
\bibitem{Lam} J. Lambe and R. C. Jaklevic, Molecular Vibration Spectra by Inelastic Electron Tunneling, {\it Phys. Rev.} {\bf 165}, 821, 1968.
\bibitem{Adk} C. J. Adkins and W. A. Phillips, Frequency shifts in inelastic electron tunnelling spectroscopy of adsorbed species, {\it J. Phys. C} {\bf 18}, 1313, 1985.
\bibitem{Tur2} L. Turin, A spectroscopic mechanism for primary olfactory reception, {\it Chem. Senses} {\bf 21}, 773, 1996.
\bibitem{Bro} J. C. Brooks {\it et al.} Could humans recognize odor by phonon assisted tunneling?, {\it Phys. Rev. Lett.} {\bf 98}, 038101, 2007.
\bibitem{Sol} I. A. Solov'yov, P.-Y. Chang and K. Schulten, Vibrationally assisted electron transfer mechanism of olfaction: myth or reality?, {\it Phys. Chem. Chem. Phys.} {\bf 14}, 13861, 2012.
\bibitem{Bit} E. R. Bittner {\it et al.} Quantum Origins of molecular recognition and olfaction in drosophila, {\it J. Chem. Phys.} {\bf 137}, 22A551, 2012.
\bibitem{Che} A. Ch\c{e}ci\'{n}ska {\it et al.}, Dissipation enhanced vibrational sensing in an olfactory molecular switch, {\it J. Chem. Phys.} {\bf 142}, 025102, 2015.
\bibitem{Tir} A. Tirandaz, F. Taher Ghahramani and A. Shafiee, Dissipative vibrational model for chiral recognition in olfaction, {\it Phys. Rev. E} {\it 92}, 032724, 2015.
\bibitem{Fra} M. I. Franco {\it et al.} Molecular vibration-sensing component in Drosophila melanogaster olfaction, {\it Proc. Natl. Acad. Sci. USA} {\bf 108}, 3797, 2011.
\bibitem{Bitt} E. R. Bittner, A. Madalan, A. Czader and G. Roman, Quantum origins of molecular recognition and olfaction in drosophila, {\it J. Chem. Phys.} {\bf 137}, 22A551, 2012.
\bibitem{Gro} W. Gronenberg {\it et al.} Honeybees (Apis mellifera) learn to discriminate the smell of organic compounds from their respective deuterated isotopomers, {\it Proc. Biol. Sci.} {\bf 281}, 20133089, 2014.
\bibitem{Haf} L. J. W. Haffenden, V. A. Yaylayan and J. Fortin, Investigation of vibrational theory of olfaction with variously labelled benzaldehydes, {\it Food Chem.} {\bf 73}, 67, 2001.
\bibitem{Gan} S. Gane {\it et al.} Molecular Vibration-Sensing Component in Human Olfaction, {\it PLoS One} {\bf 8}, e55780, 2013.
\bibitem{Kel} A. Keller and L. B.Vosshall, A psychophysical test of the vibration theory of olfaction, {\it Nat. Neurosci.} {\bf 7}, 337, 2004.
\bibitem{Blo} E. Block {\it et al.} Implausibility of the vibrational theory of olfaction, {\it Proc. Natl. Acad. Sci.} {\bf 112}, E2766, 2015.
\bibitem{Tow} C. H. Townes and A. L. Schawlow, {\it Microwave Spectroscopy}, McGraw-Hill, New York, 1955.
\bibitem{Her} G. Herzberg, {\it Molecular Spectra and Molecular Structure: Electronic Spectra and Electronic Structure of Polyatomic Molecules}, Krieger, Malabar, 1991.
\bibitem{Leg} A. J. Leggett {\it et al.} Dynamics of Dissipative two state Systems, {\it Rev. Mod. Phys.} {\bf 59}, 1, 1987.
\bibitem{Wei} U. Weiss, {\it Quantum Dissipative Systems}, World Scientific, Singapore, 2008.
\bibitem{Qua} M. Quack, How important is parity violation for molecular and biomolecular chirality?, {\it Angew. Chem. Intl. Ed.} {\bf 41}, 4618, 2002.
\bibitem{Lee} T. D. Lee and C. N. Yang, Question of Parity Conservation in Weak Interactions, {\it Phys. Rev.} {\bf 104}, 254, 1956.
\bibitem{Bar} L. D. Barron, Chirality at the sub-molecular level: True and false chirality, in {\it Chirality in Natural and Applied Science}, edited by W. J. Lough and I. W. Wainer, Blackwell Publishing, Oxford, 2002.
\bibitem{Gil} J. Gilmore and R. H. McKenzie, Spin boson models for quantum decoherence of electronic excitations of biomolecules and quantum dots in a solvent, J. Phys.: Condens. Matter {\bf 17}, 1735 (2005).
\bibitem{Tir2} A. Tirandaz, F. Taher Ghahramani and A. Shafiee, Emergence of molecular chirality due to chiral interactions in a biological environment, {\it J. Bio. Phys.} {\bf 40}, 369, 2014.
\bibitem{QuSt} M. Quack, J. Stohner and M. Willeke, High-resolution spectroscopic studies and theory of parity violation in chiral molecules, {\it Annu. Rev. Phys. Chem.} {\bf 59}, 741, 2008.
\bibitem{Ble} B. Bleaney and J. H. Loubster, Collision broadening of the ammonia inversion spectrum at high pressures, {\it Nature} {\bf 161} 522, 1948.
\bibitem{Jon} G. Jona-Lasinio, C. Presilla and C. Toninelli, Interaction induced localization in a gas of pyramidal molecules, {\it Phys. Rev. Lett} {\bf 88}, 123001, 2002.
\bibitem{Wea} R. C. Weast {\it et al.} {\it CRC handbook of chemistry and physics}, Boca Raton, FL: CRC press, 1988.
\bibitem{Bre} K. Brettel and M. Byrdin, Reaction mechanisms of DNA photolyase, {\it Curr. Opin. Struct. Biol.} {\bf 20}, 693, 2010.
\end{thebibliography}
\end{document}